# Unsnarling the Red Tape

## Computational Infrastructure for Regulatory Systems

**July 2026**

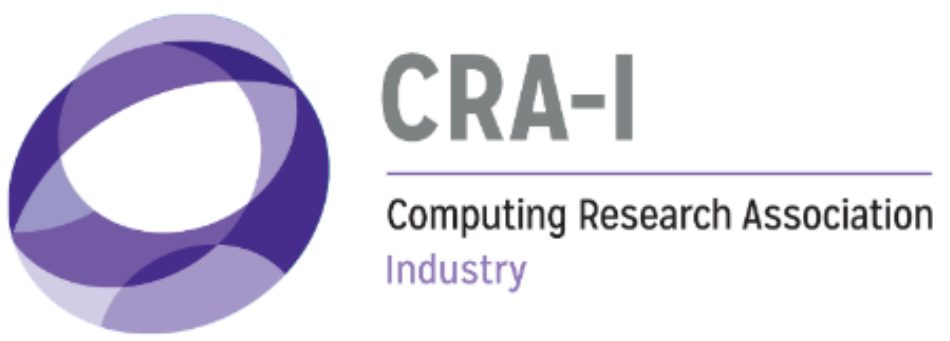

## Authors

Vinay K. Chaudhri, Knowledge Systems Research LLC
Henry F. Korth, Lehigh University
Patrick A. McLaughlin, Hoover Institution, Stanford University
Leora Morgenstern, SRI International
Jaromir Savelka, Carnegie Mellon University
Wee Kee Toh, JPMorganChase & Co.
Helen Wright, Computing Research Association

## Abstract

Regulatory complexity is increasingly recognized as an impediment to innovation, institutional responsiveness, and long-run economic growth, with regulatory accumulation estimated to reduce the U.S. GDP growth rate by nearly a full percentage point annually. This paper argues that computational approaches—including knowledge representation, artificial intelligence, natural-language processing, and cryptography—can help reduce forms of regulatory "red tape" by improving efficiency, transparency, and institutional responsiveness. We develop a framework for understanding the sources of regulatory friction and how they can be mitigated to support compliance, analysis, and reform. We further argue that effective modernization requires treating regulation not merely as legal text, but as a complex institutional and informational system that can be partially represented, analyzed, coordinated, and improved computationally while preserving legal legitimacy and procedural accountability.

## Suggested Citation

### About CRA-Industry (CRA-I)

A standing committee of the Computing Research Association (CRA), CRA-Industry (CRA-I) convenes industry partners on computing research topics of mutual interest and connects them with CRA's academic and government constituents to advance shared goals and improve societal outcomes. Through these conversations, CRA-I identifies emerging trends, develops best practices, and produces whitepapers and reports that strengthen the computing research ecosystem and drive innovation, benefiting industry and society alike.

# TABLE OF CONTENTS

# 1. INTRODUCTION

The Federal Register Act of 1935 sought to improve public access to federal rules and related executive-branch documents by establishing a centralized, authoritative, and uniform system for their filing and publication. The modern Federal Register now exceeds 100,000 pages annually (Crews 2024). Beyond federal regulations, firms, regulators, and citizens must also navigate layers of state, local, contractual, and organizational requirements that often interact in ways that are difficult to understand, coordinate, and reconcile. The resulting burdens extend beyond direct compliance costs. Regulatory accumulation can impose systemic economic and institutional costs that affect investment, innovation, productivity, and long-run economic growth.

Many of these challenges arise not solely from the existence of rules themselves, but from the difficulty of representing, analyzing, and operationalizing large interconnected systems of obligations, conditions, exceptions, approvals, and institutional procedures. While not all aspects of regulation are amenable to computation or automation, many operational components already possess implicit structure that can potentially be represented more explicitly and analyzed computationally. Advances in artificial intelligence, knowledge representation, cryptography, and natural-language processing create new possibilities for extracting such structure from existing regulations, embedding greater structural clarity into regulatory drafting, partially automating compliance workflows, improving institutional coordination, and supporting more systematic approaches to regulatory reform.

Our approach treats natural-language regulation as the authoritative source of legal meaning and computational representations as supporting artifacts rather than substitutes for legal interpretation. We therefore focus not on fully automated “law-as-code” systems, but on the broader challenge of representing, analyzing, coordinating, and operationalizing regulatory systems while preserving the central role of legal interpretation, institutional judgment, and procedural accountability. From this perspective, regulatory systems can be viewed as partially computable institutional infrastructures whose effectiveness depends on the clarity, interoperability, analyzability, and operationalizability of the rules and workflows through which they are implemented. Computational methods can improve transparency and efficiency without displacing legal interpretation or institutional judgment.

We begin by examining the sources and economic consequences of regulatory complexity. We then develop a taxonomy of regulatory frictions and discuss how different computational and institutional approaches may help mitigate them. Building on this framework, we examine structured regulatory representations, executable rule systems, institutional and workflow redesign, and computationally informed approaches to regulatory reform. Throughout, we emphasize that effective modernization requires not only technical advances, but also governance mechanisms and institutional processes capable of integrating computational systems into real-world regulatory environments responsibly and effectively.

## 2. RELATED WORK

A substantial body of scholarship has explored how knowledge representation, artificial intelligence, and natural-language processing can support regulatory compliance and reform. Efforts to formally represent legal and regulatory systems computationally are often described under the broader heading of computational law (Genesereth 2021). One of the earliest and most influential examples was the work of Robert Kowalski and colleagues at Imperial College London, who represented portions of the British Nationality Act as executable logical rules (Sergot et al., 1986). Since then, researchers in artificial intelligence and legal informatics have explored many of the challenges associated with translating regulations into computational representations, including the treatment of exceptions, discretionary judgment, legal interpretation, and procedural traceability.

More recent efforts have begun applying these ideas within operational regulatory systems. In the European Union, the GovTech4All initiative is developing a “Personal Regulation Assistant” based on structured regulatory representations and automated reasoning (European Commission). In the United States, Symbium has encoded portions of California permitting regulations to automate residential solar permitting workflows (Bakersfield Now Staff 2025). Similar approaches are increasingly visible in financial compliance, digital identity systems, and administrative workflow automation. In the banking industry, for example, Project Mandala explores a “compliance-by-design” architecture for cross-border financial transactions (Bank for International Settlements Innovation Hub 2024). The system combines structured regulatory representations, automated compliance verification, and cryptographic proof mechanisms to support compliance with well-specified requirements such as sanctions screening and capital-flow-management measures. The project provides a concrete example of how computational approaches can support scalable regulatory compliance without replacing legal interpretation or institutional oversight.

DARPA's Human-AI Communications for Deontic Reasoning DevOps (CODORD) program (DARPA 2024) ] seeks to develop methods for automatically translating obligations, permissions, and prohibitions expressed in natural language into machine-processable logical representations. The program is motivated by the longstanding knowledge-acquisition bottleneck associated with manually encoding deontic knowledge and explicitly identifies applications involving regulations, laws, policies, contracts, and compliance reasoning. CODORD complements the themes of this paper by focusing on the representation of normative knowledge and the automated construction of logical models that can support regulatory analysis and compliance reasoning.

Another important line of work focuses on computational regulatory analysis at scale. One of the most prominent examples is the RegData project, which transformed the U.S. Code of Federal Regulations into a structured dataset measuring regulatory volume, restrictiveness, industry relevance, and change over time (Al-Ubaydli and McLaughlin, 2012). Related efforts such as RegGenome (U. of Cambridge 2026) have attempted to move beyond simple text counting by extracting semantically meaningful components of regulation, including obligations, permissions, prohibitions, and institutional actors across jurisdictions. Together, these approaches demonstrate the value of treating regulatory text as analyzable data while also pointing toward richer forms of analysis capable of representing relationships within regulatory systems.

## 3. SOURCES OF REGULATORY COMPLEXITY

Regulatory complexity arises from multiple sources and cannot be attributed solely to the number of rules in a legal system. At a minimum, it arises from four distinct sources: the accumulation of rules over time, inconsistencies and defects within legal and regulatory texts, institutional incentives that sustain existing requirements, and administrative structures that distribute authority and responsibility across multiple organizations. In this section, we examine each of these sources in more detail.

### 3.1 Regulatory Accumulation as a Stock Problem

Regulatory complexity is largely a product of accumulation rather than design. Rules are layered onto existing rules, and they are rarely revisited, simplified, or systematically pruned. The result is a dense and opaque regulatory stock that firms, workers, consumers, and regulators must navigate in its entirety. From an economic perspective, this is a stock problem rather than merely a problem of interpreting individual rules. Each new regulation may appear modest in isolation, but economic actors make decisions against the full accumulated set of requirements. In this respect, regulatory accumulation resembles technical debt: individual rules may solve local problems, but over time, the interactions, exceptions, cross-references, and maintenance costs of the overall system can become prone to inconsistency, difficult to manage, and resistant to modernization.

### 3.2 Structural Defects in Legal and Regulatory Texts

Regulatory complexity arises not only from the accumulation of rules, but also from defects embedded within individual legal and regulatory instruments. Such defects may take the form of contradictory requirements or category errors. As an example of a contradictory requirement, the Utah Supreme Court found a statutory provision "entirely inoperable" because it contained irreconcilable commands (*Nelson v. Salt Lake County,* 1995). As an example of a category error, the Supreme Court observed that a provision of the Indian Gaming Regulatory Act identified tax-reporting and withholding provisions as applicable to Indian gaming operations but then cited a chapter of the Internal Revenue Code governing wagering taxes rather than reporting requirements (*Chickasaw Nation v. United States,* 2001). These defects arise independently of regulatory accumulation, although they are often compounded as legal texts are amended, interpreted, and integrated with other statutes and regulations over time.

### 3.3 Institutional Sources of the Persistence of Regulatory Complexity

Institutional incentives also contribute to why regulatory complexity persists. Agencies are often punished more severely for visible errors of permission—allowing an activity that later causes harm—than for errors of prohibition, such as delayed projects, foregone entry, abandoned innovations, or dispersed compliance costs borne by firms and consumers. This asymmetry can produce precautionary inertia in which rules are more likely to be added than removed, even when underlying technologies, markets, or operational conditions have changed substantially.

### 3.4 Administrative and Coordination Complexity

Regulatory complexity arises not only from the substance of rules themselves, but also from the administrative processes through which those rules are implemented and enforced. In many domains, firms and individuals must navigate overlapping approvals, fragmented reporting requirements, and interagency dependencies. Even when individual rules are justified on their own terms, the cumulative interaction of procedural requirements and fragmented oversight structures can generate substantial delays, uncertainty, and operational costs.

## 4. CONSEQUENCES OF REGULATORY COMPLEXITY

Regulatory complexity generates multiple forms of friction that affect firms, regulators, workers, and consumers. These frictions can be broadly categorized as informational, procedural, organizational, and dynamic.

### 4.1 Informational Friction

One class of friction is informational. Firms must first determine which rules apply to their activities, jurisdictions, products, or organizational structures. As regulatory systems grow, this search problem becomes increasingly difficult because relevant requirements are distributed across statutes, agency rules, guidance documents, standards, contracts, and organizational policies. Even after relevant rules are identified, substantial interpretation frictions remain. Firms and agencies must resolve ambiguity, cross-references, exceptions, conflicting provisions, and evolving interpretations. These informational burdens can create uncertainty and increase the costs of planning, investment, and operational decision-making.

### 4.2 Procedural Friction

The second class of friction is procedural. Many regulatory systems involve sequential approvals, fragmented reporting requirements, interagency coordination, licensing procedures, audits, and recurring documentation obligations. In such settings, the burden often arises not simply from the existence of a rule, but from the fragmented workflows surrounding its implementation and enforcement. Delay costs are especially important in permitting-and-licensing systems, where projects, capital, or labor may remain idle while approvals are pending.

Infrastructure permitting presents a vivid example of procedural friction. Major projects often require coordination across multiple agencies with overlapping jurisdictions and procedural requirements, creating delays and uncertainty that can persist for years (Bull and McLaughlin, 2026).

### 4.3 Organizational Friction

A third class of friction is organizational and arises from differences in institutional capacity, scale, and administrative resources. Compliance systems frequently impose fixed costs that are disproportionately difficult for small firms and new entrants to absorb. Large incumbent firms may be able to distribute compliance costs across greater scale, maintain specialized legal and regulatory staff, or adapt more easily to procedural complexity.

### 4.4 Dynamic Friction

Finally, regulatory complexity generates dynamic frictions that impede adaptation and change over time. Accumulated rules and procedural uncertainty can alter investment decisions, discourage experimentation, delay technological adoption, and reduce incentives for market entry. The long-run consequences can include lower productivity growth, slower diffusion of beneficial technologies, and reduced economic dynamism. Transportation provides a particularly clear example of dynamic regulatory friction. In freight rail, automated track inspection systems can improve both safety and operational efficiency, yet regulatory barriers have slowed their adoption despite demonstrated benefits (Ellig and McLaughlin, 2016; Coffey and McLaughlin 2026a; Coffey and McLaughlin, 2026b).

At an aggregate level, these dynamic frictions can have measurable economic consequences. Empirical work estimates that cumulative federal regulation slowed annual U.S. GDP growth by roughly 0.8 percentage points since 1980, primarily by distorting investment decisions and reducing long-run capital formation (Coffey et al., 2020). Had regulation remained at 1980 levels, their model predicts that the U.S. economy would have been nearly 25 percent larger by 2012, equivalent to roughly $4 trillion in lost output, or approximately $47,000 per household in current dollars. A broader literature review finds related effects on entrepreneurship (Bailey and Thomas, 2017), employment growth (Bailey and Thomas, 2017), income inequality (Chambers et al. 2019a; Chambers and O'Reilly, 2022), poverty (Chambers et al., 2019b), and consumer prices (Chambers et al., 2019), suggesting that regulatory accumulation can disproportionately burden small firms, new entrants, lower-income households, and innovation-intensive sectors (Bailey and Thomas, 2017). These findings do not imply that regulation is inherently undesirable, but rather that the accumulated stock of rules can impose systemic costs that are often difficult to observe through conventional compliance-cost measures alone.

## 5. COMPUTER SCIENCE AGENDA TO TAME REGULATION COMPLEXITY

Not every source of regulatory complexity is amenable to computational solutions, but some are. Any such exploration must begin by asking a fundamental question: Is it the case that the regulation itself needs reform? Or, is it the case that the regulation is well thought out, and we need more efficient ways to comply with it? A blind application of computational tools may entrench inefficiencies rather than reduce them. Hence, the agenda must address both regulatory compliance and regulatory reform. We should look for ways to reduce the cost of complying with rules that serve legitimate public purposes, and at the same time, support reform by improving the ability of policymakers, researchers, and the public to identify rules that are obsolete, duplicative, conflicting, or unnecessarily costly.

The taxonomy of regulatory frictions introduced in the previous section suggests that each requires different computational and institutional responses. Broadly speaking, we consider three complementary approaches: structured regulatory representations, executable compliance rule systems, and institutional and workflow redesign.
Structured regulatory representations aim to make regulatory systems more explicit and analyzable by identifying and organizing elements such as obligations, permissions, conditions, exceptions, workflows, dependencies, and reporting requirements. These representations do not necessarily automate compliance. Instead, their primary purpose is to improve navigation and interpretation of regulations, and large-scale regulatory analysis. We consider this approach further in Section 6.

Executable compliance rules operate at a more detailed operational level. They encode sufficiently precise regulatory and operational requirements as computer-executable logic capable of supporting compliance decisions, transaction processing, reporting, approvals, screening, or enforcing workflows. Such systems are most applicable in domains where requirements are well specified, operationally measurable, and amenable to deterministic evaluation. We consider such systems in Section 7.
A third approach involves institutional and workflow redesign. Effective deployment of computational systems may require new governance processes, coordination mechanisms, and administrative workflows capable of integrating structured representations and executable rule systems into existing institutional environments. We will consider this approach in greater detail in Section 8.

In general, each computational approach can help mitigate multiple forms of regulatory friction and can support both regulatory compliance and regulatory reform. Table 1 illustrates these overlapping relationships between computational interventions and the primary forms of regulatory friction they are intended to address. In the subsequent sections, we consider each of these approaches, and then also discuss how they can support a systematic analysis to support regulatory reform.

| | Informational | Procedural | Organizational | Dynamic |
|---|---|---|---|---|
| Structured regulatory representations | x | x | | x |
| Executable compliance rule systems | x | x | | x |
| Institutional and workflow redesign | | x | x | x |

Table 1: Computational interventions and their direct influence on mitigating regulatory friction

## 6. STRUCTURED REPRESENTATION FOR REGULATION

We can develop structured representations of regulation through two complementary strategies: extracting structure from existing regulatory text using computational methods, and embedding greater structural clarity into regulations during its initial drafting. The objective is not complete formalization of regulation, but increasing the degree of structure available for navigation, and analysis. We consider these two approaches in detail.

## 6.1 Computational Extraction of Regulatory Structure

Recent advances in natural-language processing (NLP) technologies and large language models (LLMs) suggest new possibilities for extracting structured representations from legal and regulatory text. LLMs and even simpler NLP technologies can assist in identifying regulated entities, obligations, conditions, exceptions, and decision pathways embedded within legislative language. Initial work demonstrates that such systems can generate candidate decision trees and executable logical rules from statutes and regulations (Janatian et al., 2023).

Earlier LLMs required extensive expert validation, but recent advances in model capabilities, agentic workflows, and reinforcement-learning-based alignment techniques have significantly improved the reliability of extracting structured information from complex regulatory texts. Growing investment in dedicated research programs focused on translating natural-language regulations into formal representations further suggests that portions of this knowledge-engineering process may be amenable to automation (Morgenstern, 2026). Current systems such as RegData transform regulatory corpora into datasets measuring regulatory volume, restrictiveness, industry relevance, and change over time (Al-Ubaydli and McLaughlin, 2017). The next generation should move beyond counting restrictive terms toward extracting the semantic structure of regulation itself: who is regulated, what actions are required or prohibited, under what conditions, subject to which exceptions, at what time, and in relation to which other rules. In economic terms, the objective is to move from measuring the size of the regulatory stock to measuring its structure.

Future progress will require a shared scientific infrastructure for computational regulatory analysis, including benchmark datasets, expert-annotated corpora, and standardized evaluation frameworks spanning multiple regulatory domains. Such infrastructure would enable systematic comparison of methods for extracting, representing, and analyzing obligations, exceptions, dependencies, and conflicts, while supporting more rigorous measurement of regulatory complexity, accumulation, and evolution over time.

## 6.2 Embedding Greater Structural Clarity into Regulations

While one approach to computational regulatory analysis involves extracting structure from existing legal text, a complementary approach is to embed greater structural clarity into regulations during drafting itself. Current regulations are often written primarily for human interpretation, with limited standardization in the representation of obligations, exceptions, temporal conditions, workflows, dependencies, and cross-references. As a result, even relatively straightforward compliance requirements can become difficult to interpret consistently across institutions and jurisdictions. Increasing the degree of explicit structure within regulatory drafting could improve clarity for both human and computational users while reducing ambiguity, inconsistency, and implementation costs.

Achieving this objective would require more regularized representational conventions for regulatory drafting. Earlier work in artificial intelligence and knowledge representation explored similar challenges in commonsense reasoning and the formal representation of complex domains. Researchers such as John McCarthy (McCarthy, 1986), Ernest Davis (Davis, 2014), and Murray Shanahan (Shanahan, 1997) developed methods for systematically representing predicates, functions, defaults, exceptions, and evolving operational states. These ideas are especially relevant for regulatory systems because many regulations are fundamentally structured around layered obligations and exceptions. For example, anti–money laundering (AML) compliance may require all transactions to be screened against sanctions lists, while allowing simplified procedures for low-risk customers below specified thresholds, yet still requiring enhanced due diligence for transactions involving high-risk jurisdictions or counterparties. Structurally, such systems involve nested obligations, exception handling, temporal conditions, workflows, and state transitions that are difficult to represent clearly using purely narrative drafting conventions.

More explicit structural conventions could help represent these relationships directly rather than leaving them implicit within natural-language text. Regulatory drafting could incorporate clearer representations of obligations, exceptions, dependencies, workflows, and operational state changes while preserving the flexibility and interpretive richness of natural language. Over time, this will increase the degree of accessible and reusable structure embedded within regulatory systems themselves.

## 7. EXECUTABLE COMPLIANCE-RULE SYSTEMS

Compliance automation is naturally suited for regulations that do not require any human involvement for interpretation, and/or external measurements. We consider two concrete gaps, which, if addressed, can substantially enhance our ability to build automated compliance systems: repositories of executable rules and privacy-preserving compliance.

### 7.1 Executable Compliance-Rule Repositories

Effective compliance automation depends on the ability to represent sufficiently precise regulatory and operational requirements as executable decision logic over data, enabling determinate outcomes such as approval, denial, reporting, escalation, or transaction execution.

Yet relatively few publicly accessible repositories of reusable executable compliance rules currently exist to support interoperability, standardization, auditability, and scalable automation across institutions.

The relevant rules are not limited to externally imposed regulations. A common example arises in commercial banking through zero-balance account (ZBA) structures used in corporate treasury management (Toh et al., 2024). In such systems, operational accounts are maintained at or near zero-balance by automatically transferring funds to or from a central master account. Incoming funds are swept into the master account, while outgoing payments trigger just-in-time funding to satisfy obligations. These processes can naturally be represented as executable trigger–condition–action rules in which events such as payment initiation or end-of-day processing trigger balance checks and corresponding transfers.

Similar patterns arise across domains such as engineering standards, permitting, environmental review, and transportation, where rules are often operational, economically consequential, and tied to measurable outcomes (Montgomery et al., 2019; The White House, 2025; Coffey and McLaughlin, 2026a). These characteristics make such domains especially amenable to structured computational representation and scalable compliance automation. Shared repositories of regulatory and operational rules could improve consistency, auditability, reuse, and interoperability across institutions while reducing the repeated cost of translating regulatory requirements into operational systems.

### 7.2 Privacy-Preserving Compliance Automation

Without adequate privacy protections, large-scale compliance automation risks creating new forms of surveillance, data concentration, cybersecurity vulnerability, and loss of institutional trust. Compliance automation often requires access to highly sensitive information, including financial transactions, medical records, personal identities, and proprietary business information. Conventional compliance architectures frequently rely on centralized data collection, broad information disclosure, retrospective auditing, and after-the-fact detection of violations, requiring organizations to reveal substantially more information than is necessary to establish compliance with a particular rule.

Recent advances in cryptography provide an alternative foundation for privacy-preserving compliance automation. Once regulatory requirements are represented in sufficiently structured computational form, cryptographic techniques such as zero-knowledge proof can enable compliance verification while revealing only the minimum information necessary for validation (Goldwasser et al., 1985). Instead of disclosing underlying transactional data directly, a system can generate proofs that a transaction satisfies particular regulatory constraints without exposing sensitive details such as transaction amounts, counterparties, account balances, medical records, or proprietary business logic (Duff, 2026).

Such approaches are particularly suited to well-specified and machine-checkable constraints, including sanctions screening, eligibility requirements, reporting thresholds, and other deterministic compliance conditions, though they are less applicable in domains requiring substantial legal interpretation or discretionary judgment. For example, anti-money-laundering and sanctions requirements are commonly implemented through checks against restricted-party databases. Emerging systems demonstrate how cryptographic proofs can support compliance with membership and non-membership requirements while preserving transactional privacy (Railgun, 2024).

Future compliance infrastructures should therefore integrate cryptographic methods more directly into rule-based compliance architectures. Doing so would allow institutions to demonstrate compliance while minimizing unnecessary disclosure of sensitive information, enabling forms of real-time verification that are both more privacy-preserving and more operationally scalable. Privacy preservation should therefore be treated not as an optional feature of compliance automation, but as a foundational requirement for trustworthy computational regulatory systems.

## 8. INSTITUTIONAL AND WORKFLOW DESIGN

Computational approaches alone are unlikely to substantially reduce regulatory friction without corresponding changes to institutional processes and administrative workflows. As computational regulatory systems become more capable, institutions will need to adapt their workflows, governance mechanisms, and reform processes in ways that allow these technologies to be deployed effectively.

### 8.1 Improving Administrative Efficiency

One important opportunity lies in reducing procedural and administrative delays. Delay costs can often be reduced through automated eligibility checks, permitting-status transparency, standardized workflows, and, where legally authorized, shot clocks that trigger default approval, escalation, or mandatory review after a predetermined period of agency inaction. While computational systems can support automated eligibility checks and workflow coordination, mechanisms such as default approval, escalation, and mandatory review ultimately require institutional and legal redesign. Likewise, permitting-status transparency and standardized workflows often depend less on technological feasibility than institutional willingness to adopt existing tools for these purposes.

### 8.2 Institutional Support for Structured Regulatory Systems

Beyond technical feasibility, large-scale adoption of computational regulatory systems raises important questions concerning governance, incentives, and institutional authority. A practical approach is to treat structured and machine-readable regulatory representations as complementary artifacts to existing legal text rather than as replacements for it. Under such a model, agencies could incrementally develop and publish structured versions of high-impact regulations while preserving the legal primacy of natural-language statutes and rules.

Ownership and maintenance of these representations could remain distributed, with domain agencies responsible for maintaining authoritative rule definitions supported by shared standards and infrastructure developed collaboratively across government, academia, and industry. Public-private partnerships and open repositories may help accelerate adoption, while expert validation workflows remain essential. Over time, as these systems demonstrate value in reducing compliance costs, improving transparency, and supporting more consistent enforcement, they may become increasingly integrated into ordinary regulatory processes, much as digital filing-and-disclosure systems have today.

## 9. COMPUTATIONALLY INFORMED REGULATORY DESIGN

The preceding sections focused primarily on how computational systems can support compliance automation, structured regulatory representations, and improved administrative workflows. A further possibility is that these same systems may also support more systematic approaches to regulatory analysis and reform.

### 9.1 Using Structured Regulations to Support Reform

Structured representations of obligations, conditions, exceptions, workflows, and cross-references would allow researchers and policymakers to distinguish regulations that are routine from those that are ambiguous, duplicative, highly consequential, or innovation-constraining. Applied longitudinally, such methods could reveal patterns of regulatory accumulation, fragmentation, and procedural complexity over time, creating the foundation for more systematic and data-driven approaches to regulatory reform.

Recent state-level reforms suggest that these approaches are beginning to move from theory into practice. Virginia's 2025 Executive Order 51 launched an agentic AI pilot program to analyze regulations and guidance documents for contradictions, redundancies, simplification opportunities, and cross-reference inconsistencies (Bull and McLaughlin, 2026). Although much of the reported reduction in regulatory requirements preceded deployment of the AI system itself, the pilot illustrates how computational tools may support scalable regulatory review and prioritization for expert analysis.

### 9.2 Regulatory Inventories and Budgeting

Computationally-structured rule repositories may also support more systematic forms of regulatory inventory management and regulatory budgeting. Regulatory budgeting requires visibility into what requirements exist, which agencies produced them, whom they affect, and how they change over time. Without such inventories, policymakers face substantial difficulty managing the growth and accumulation of regulation across agencies and sectors.
British Columbia's regulatory reform experience illustrates the importance of maintaining transparent inventories of regulatory requirements. The province reduced regulatory requirements substantially relative to its 2001 baseline while maintaining a public accounting system that made additions and reductions visible. These reforms were subsequently associated with measurable increases in economic growth (Coffey and McLaughlin, 2021). Structured regulatory repositories could provide the informational infrastructure necessary to support similar forms of ongoing regulatory measurement, transparency, and budgeting in other jurisdictions.

### 9.3 Dynamic Analysis of Regulatory Accumulation

More broadly, computational representations may allow researchers and policymakers to study regulation as an evolving institutional system. Structured representations of obligations, exceptions, workflows, and cross-references create opportunities for longitudinal analysis of how regulatory systems grow, fragment, and interact over time.

Such capabilities may help identify regulations that are obsolete, duplicative, highly interconnected, innovation-constraining, or disproportionately costly relative to their intended benefits. Embedding-based clustering, semantic-similarity analysis, and rule-extraction methods may also help reveal hidden relationships across agencies and regulatory domains that would be difficult to detect through manual analysis alone. Over time, these approaches could support more evidence-based and data-driven approaches to regulatory modernization, allowing policymakers to better distinguish between regulation that is socially beneficial, regulation that is administratively inefficient, and regulation whose costs may substantially exceed its benefits.

## 10. CONCLUSION

Modern regulatory systems have become increasingly complex due to decades of incremental accumulation, resulting in significant economic costs, slower innovation, administrative delay, and reduced transparency. We have proposed a multi-layered approach to regulatory modernization combining compliance automation, greater structural clarity in regulatory systems, institutional and workflow redesign, and computationally informed regulatory reform. Any such system must ultimately be evaluated not only on computational performance, but also on whether it meaningfully reduces administrative "red tape" through lower compliance costs, reduced delay, and greater transparency while preserving legal fidelity, institutional legitimacy, and procedural fairness. Ultimately, the goal is not to replace human judgment or legal interpretation, but to treat regulation as a complex institutional system that can be partially represented, analyzed, coordinated, and systematically improved through computational methods.

## ACKNOWLEDGMENT

This work has been supported by the National Science Foundation under Award No 2514820. Any opinions, findings and conclusions or recommendations expressed in this material are those of the author(s) and do not necessarily reflect the views of the National Science Foundation. This paper resulted from a roundtable hosted by the CRA-Industry Committee. The topic of discussion in the round table was *Unsnarling the Red Tape: Machine-Interpretable Rule Systems for Efficient Regulatory Compliance and Reform.* We thank Prof. Cogan Shimizu for his feedback on the report.

---